\documentclass[prl,aps,superscriptaddress,preprintnumbers,twocolumn,notitlepage,nofootinbib]{revtex4-1}
\usepackage{amsmath,amssymb,graphicx,bm}
\usepackage{enumerate}

\newcommand{\beq}{\begin {equation}}  
\newcommand{\eeq}{\end   {equation}} 
\newcommand{\bea}{\begin {eqnarray}} 
\newcommand{\eea}{\end   {eqnarray}}  
\newcommand{\baa}{\begin {array}   } 
\newcommand{\eaa}{\end   {array}   }     
\newcommand{\bit}{\begin {itemize} }
\newcommand{\eit}{\end   {itemize} }
\newcommand{\be }{\begin {equation}} 
\newcommand{\ee }{\end   {equation}}
\newcommand{\nn }{\nonumber        }

\allowdisplaybreaks
\begin{document}


\preprint{ACFI-T16-19}

\title{The Radiative $\mathbb{Z}_2$ Breaking Twin Higgs}

\author{Jiang-Hao Yu} 
\affiliation{Amherst Center for Fundamental Interactions, Department of Physics, University of Massachusetts-Amherst, Amherst, MA 01003, U.S.A.}


\begin{abstract}

In twin Higgs model, the Higgs boson mass is protected by a $\mathbb{Z}_2$ symmetry.  
The $\mathbb{Z}_2$ symmetry needs to be broken either explicitly or spontaneously to obtain misalignment between electroweak and new physics vacua.  
We propose a novel $\mathbb{Z}_2$ breaking mechanism, in which the $\mathbb{Z}_2$  is spontaneously broken 
by radiative corrections to the Higgs potential.
Two twin Higgses with different vacua are needed, and vacuum misalignment is realized by
opposite but comparable contributions from gauge and Yukawa interactions to the potential. 
Due to fully radiative symmetry breaking, the Higgs sector is completely determined by twin Higgs vacuum, Yukawa and gauge couplings. 
There are eight pseudo-Goldstone bosons: the Higgs boson, inert doublet Higgs, and three twin scalars.
We show the  125 GeV Higgs mass  and constraints from Higgs coupling measurements could be satisfied.

\end{abstract}

\maketitle



The discovery of a 125 GeV Higgs boson at the LHC~\cite{Aad:2012tfa}   
sharpens existing naturalness problem 
in the Standard Model (SM): quadratically divergent quantum corrections to the Higgs boson mass destabilize the electroweak scale. 
This suggest the existence of new physics (NP) with a new symmetry which protects the Higgs mass against large radiative corrections. 
In supersymmetry and composite Higgs~\cite{Kaplan:1983fs, ArkaniHamed:2002qy}  models, SM partners from new symmetry play the role of stabilizing the Higgs mass.
Unfortunately, null results on new physics searches at the LHC put tight lower bounds on them.
This leads to a sub-percent level of tuning between electroweak and NP cutoff scales, which is the little hierarchy problem~\cite{Barbieri:2000gf}.

The twin Higgs model~\cite{Chacko:2005pe} [see also~\cite{Craig:2014aea, Burdman:2014zta, Craig:2015pha}] is introduced to address the little hierarchy problem. 
It introduces a mirror copy of the SM: the twin sector, 
which is completely neutral under the SM gauge group. 
Since twin partners are colorless, they could have sub-TeV masses and thus soften the little hierarchy. 
The approximate global symmetry breaking $U(4)/U(3)$ at scale $f$ produces a pseudo Goldstone boson (PGB), identified as the Higgs boson.   
Imposing a discrete $\mathbb{Z}_2$ symmetry between SM and twin sector ensures that there is no
quadratically divergent quantum corrections to the Higgs mass term. 
The $\mathbb{Z}_2$ symmetry needs to be broken to realize vacuum misalignment mechanism: how to generate asymmetric vacua  $v < f$ for the Higgs boson and twin Higgs boson.
In original twin Higgs model, the $\mathbb{Z}_2$ symmetry is broken explicitly by introducing soft or hard $\mathbb{Z}_2$ breaking terms
in scalar potential. 
This minimal model has been extended to incorporate two twin Higgses in non-supersymmetric~\cite{Chacko:2005vw} and supersymmetric~\cite{Chang:2006ra} frameworks.
The advantage of two twin Higgses setup is that it could accommodate a different $\mathbb{Z}_2$ breaking mechanism~\cite{Beauchesne:2015lva, Harnik:2016koz}: 
the $\mathbb{Z}_2$  symmetry is spontaneously broken by a bilinear Higgs mass term between two twin Higgses. 
The vacuum expectation value (VEV) of one twin Higgs preserves $\mathbb{Z}_2$, while the other breaks $\mathbb{Z}_2$ completely and spontaneously. 
As the effective tadpole, this bilinear term transmits the $\mathbb{Z}_2$ breaking from the broken one to the unbroken one, 
and thus obtain vacuum misalignment.

In this work, we propose a novel approach to spontaneously break the $\mathbb{Z}_2$ symmetry: the radiative $\mathbb{Z}_2$ breaking mechanism. 
The Higgs potential is fully generated from gauge and Yukawa corrections, and the $\mathbb{Z}_2$ symmetry is broken spontaneously and radiatively. 
The radiatively generated Higgs potential is parametrized as
\bea
	V(h) \simeq \frac{g_{\rm SM}^2 m_{*}^2}{16\pi^2}\left( - a |h|^2 + b \frac{|h|^4}{f^2}\right),
\eea
where $g_{\rm SM}$ is a typical SM coupling and $m_{*}$ is the mass scale of twin partner. 
Typically radiative corrections have $a$ and $b$ at the same order, which only induce symmetric vacua $v = f$. 
To realize vacuum misalignment $v < f$, we need either $a$ is suppressed or $b$ is enhanced. 
For example, in littlest Higgs~\cite{ArkaniHamed:2002qy}  the quartic term $b$ is enhanced via adding tree-level quartic terms by hand. 
Without adding terms by hand, we could utilize possibly large cancellation among radiative corrections to suppress quadratic term $a$. 
In the original twin Higgs, 
we note that gauge and Yukawa corrections to the quadratic term $a$ have opposite sign.  
However, a large cancellation cannot happen because gauge corrections are much smaller than Yukawa ones. 
Interestingly, gauge corrections can be enhanced by introducing a second twin Higgs with global symmetry breaking scale $f' \gg f$. 
This causes comparable but opposite gauge and Yukawa corrections, and leads to  vacuum misalignment with a moderate tuning between $v$ and $f$. 
Thus, a different but more minimal spontaneous $\mathbb{Z}_2$ breaking mechanism  is naturally realized without introducing either a soft breaking term or a bilinear tadpole term. 
%



Two $U(4)$ invariant Higgs fields are introduced as
\bea
	H_1 \equiv \left(\begin{array}{c} H_{1A} \\ H_{1B} \end{array}\right),\qquad H_2 \equiv \left(\begin{array}{c} H_{2A} \\ H_{2B} \end{array}\right).
\eea
where two twin Higgs doublets $H_{1B}$ and $H_{2B}$ are in twin sector. 
The $\mathbb{Z}_2$ symmetry maps the twin Higgses into visible Higgses: $H_{1B} \xrightarrow{\mathbb{Z}_2} H_{1A}$, $H_{2B} \xrightarrow{\mathbb{Z}_2} H_{2A}$.
The scalar pontential, which respects both ${\mathbb{Z}_2}$ and global $U(4)_1 \times U(4)_2$ symmetries, reads 
\bea
	V(H_1, H_2) &=& 
	-\mu_1^2 |H_1|^2 - \mu_2^2 |H_2|^2 \nn\\
	&& + \lambda_1 |H_1|^4 + \lambda_2 |H_2|^4
	+ \lambda_3 |H_1|^2 |H_2|^2.
	\label{eq:2HDMPot}
\eea
The Higgs sector is weakly gauged under both the SM and the mirror SM gauge symmetries. 
After symmetry breaking $\langle H_i \rangle \equiv f_i$ ($i=1,2$), the symmetries of the Lagrangian have
\bea
  \textrm{global   symmetry:} && \quad U(4) \times U(4) \to U(3) \times U(3),\\
  \textrm{gauge    symmetry:}  && \quad [SU(2) \times U(1)]_{A,B} \to [SU(2) \times U(1)]_A. \nn
\eea 
In nonlinear $\sigma$ Lagrangian, assuming radial modes in $H_i$ are decoupled, 
the fields $H_i$ are parametrized as
\bea
	H_i = \exp\left[\frac{i}{f_i}\left(\begin{array}{c|cc}
		\bm{0}_{2\times 2}&\bm{0}_{1 \times 2}&\bm{h}_1\\\hline
		\bm{0}_{2 \times 1}&0&C_i\\ 
		\bm{h}_i^{\ast}&C_i^{\ast}&N_i
		\end{array} \right)\right]\left(\begin{array}{c} \bm{0}_{1 \times 2} \\\hline 0 \\ f_i \end{array}\right),
\eea
where 14 GBs $\bm{h}_i, C_i, N_i (i = 1,2)$   are generated.

There are two ways to incorporate fermions. 
In the ``mirror fermion" assignment~\cite{Chacko:2005pe, Craig:2014aea}, the SM fermions have mirror fermions: 
$q_A(3,2;1,1) \xrightarrow{\mathbb{Z}_2} q_B(1,1;3,2)$ and $t_A(3,1;1,1) \xrightarrow{\mathbb{Z}_2} t_B(1,1;3,1)$,   
with quantum number assignment $[SU(3), SU(2)]_{A,B}$.
The general top-Yukawa Lagrangian reads
\bea
	- {\mathcal L}_{\rm Yuk} &=& y_1 \left(H_{1A}^\dagger q_A  \bar{t}_A + H_{1B}^\dagger q_B \bar{t}_B\right) + (1 \leftrightarrow 2) + h.c.
\eea
To avoid Higgs mediated flavor changing neutral current in $A$ sector, similar to the two Higgs doublet model (2HDM), 
either the discrete $Z'_2$ symmetry or aligned Yukawa structure~\cite{Branco:2011iw}
are imposed.
We will discuss the following two Yukawa structures: Type-I Yukawa structure ($y_2 = 0$), 
and a special aligned Yukawa structure $y_2 = \epsilon y_1$ ($\epsilon \ll 1$)~\cite{Pich:2009sp}.  
In the ``$U(4)$ fermion" assignment~\cite{Chacko:2005pe}, the following $U(4)$ fermions are introduced:
\bea
	Q &=& q_A + q_B + \tilde{q}_A \,(3,1; 1,2) + \tilde{q}_{B}\, (1,2;3,1),\nn\\
	U &=& t_A \,(3,1; 1,1) + t_B \,(1,1; 3,1). 
\eea
The top Yukawa Lagrangian is
\bea
	- {\mathcal L}_{\rm Yuk} = y_1 H_1^\dagger Q \bar{U} + y_2 H_2^\dagger Q \bar{U} + M \bar{\tilde{q}}_{A,B}\tilde{q}_{A,B}+ h.c. 
\eea
Here either Type-I or aligned Yukawa structure is used.

The global  $U(4)\times U(4)$ symmetry is weakly broken by the radiative corrections from the gauge and Yukawa interactions.  
The dominant radiative corrections to the scalar potential are written as
\bea
	V_{\rm loop} &=& \delta_1  |H_{1A}|^4  
	+ \delta_2  |H_{2A}|^4   + \delta_3  |H_{1A}|^2 |H_{2A}|^2  \nn\\
	&& + \delta_4  |H_{1A}^\dagger H_{2A}|^2     + \frac{\delta_5}{2} \left[(H_{1A}^\dagger H_{2A})^2  + h.c. \right]\nn\\
	&&
	+  \left[  (\delta_6|H_{1A}|^2 + \delta_7|H_{2A}|^2) H_{1A}^\dagger H_{2A} + h.c. \right] \nn\\
	&& +  (A \leftrightarrow B).
	\label{eq:loopcorr}
\eea
Here for gauge bosons $W_{A,B}$ and  $Z_{A,B}$, the one-loop corrections are  
\bea
	\delta_{1,2}^B &\simeq &  -\frac{1}{16 \pi^2} \left( \frac98 g^4 + \frac34 g^2 g'^2 + \frac38 g'^4 \right) \log\frac{\Lambda}{f},\nn\\
	\delta_3^B &\simeq &  -\frac{1}{16 \pi^2} \left( \frac94 g^4 -      \frac32 g^2 g'^2 + \frac34 g'^4 \right) \log\frac{\Lambda}{f},\nn\\
	\delta_4^B &\simeq &  -\frac{1}{16 \pi^2} \left(       3 g^2 g'^2  \right) \log\frac{\Lambda}{f},\quad
	\delta_{5-7}^B \equiv 0,
\eea
where $f \equiv \sqrt{f_1^2 + f_2^2}$ and $\Lambda \equiv 4\pi f$.
For fermions, radiative corrections depend on the fermion assignment and Yukawa structure. 
In the Type-I Yukawa structure, we obtain~\cite{Chacko:2005pe}
\bea
	\delta_1^F &\simeq & \begin{cases}		
\frac{3 y^4}{16 \pi^2}\log\frac{\Lambda^2}{f^2} \qquad \qquad\qquad\qquad\qquad\text{(mirror fermion)} \\
\frac{3 y^4}{16 \pi^2} \frac{y}{x(z-x)}\left[x \log\frac{z+x}{x} - (x \leftrightarrow z)\right]   \,\,\,  \text{($U(4)$ fermion)} 
\end{cases},
\eea
where $x = y_1^2 f^2$ and $z = M^2$, 
and $\delta^F_{2-7} = 0$. 
In the aligned Yukawa structure with $y_2 \ll y_1$, we have
\bea
	&&\delta_{1,2}^F \simeq  \frac{3 y^4_{1,2}}{16 \pi^2}  \log\frac{\Lambda^2}{f^2},\quad
	\delta_{3-5}^F \simeq   \frac{3 y^2_1 y^2_2}{16 \pi^2}  \log\frac{\Lambda^2}{f^2},\nn\\
	&&\delta_{6}^F \simeq \frac{3 y^3_1 y_2}{16 \pi^2}   \log\frac{\Lambda^2}{f^2},\quad
	\delta_{7}^F \simeq \frac{3 y_1 y^3_2}{16 \pi^2}   \log\frac{\Lambda^2}{f^2}.
\eea



\begin{figure}[!t]
\includegraphics[width=4.2cm]{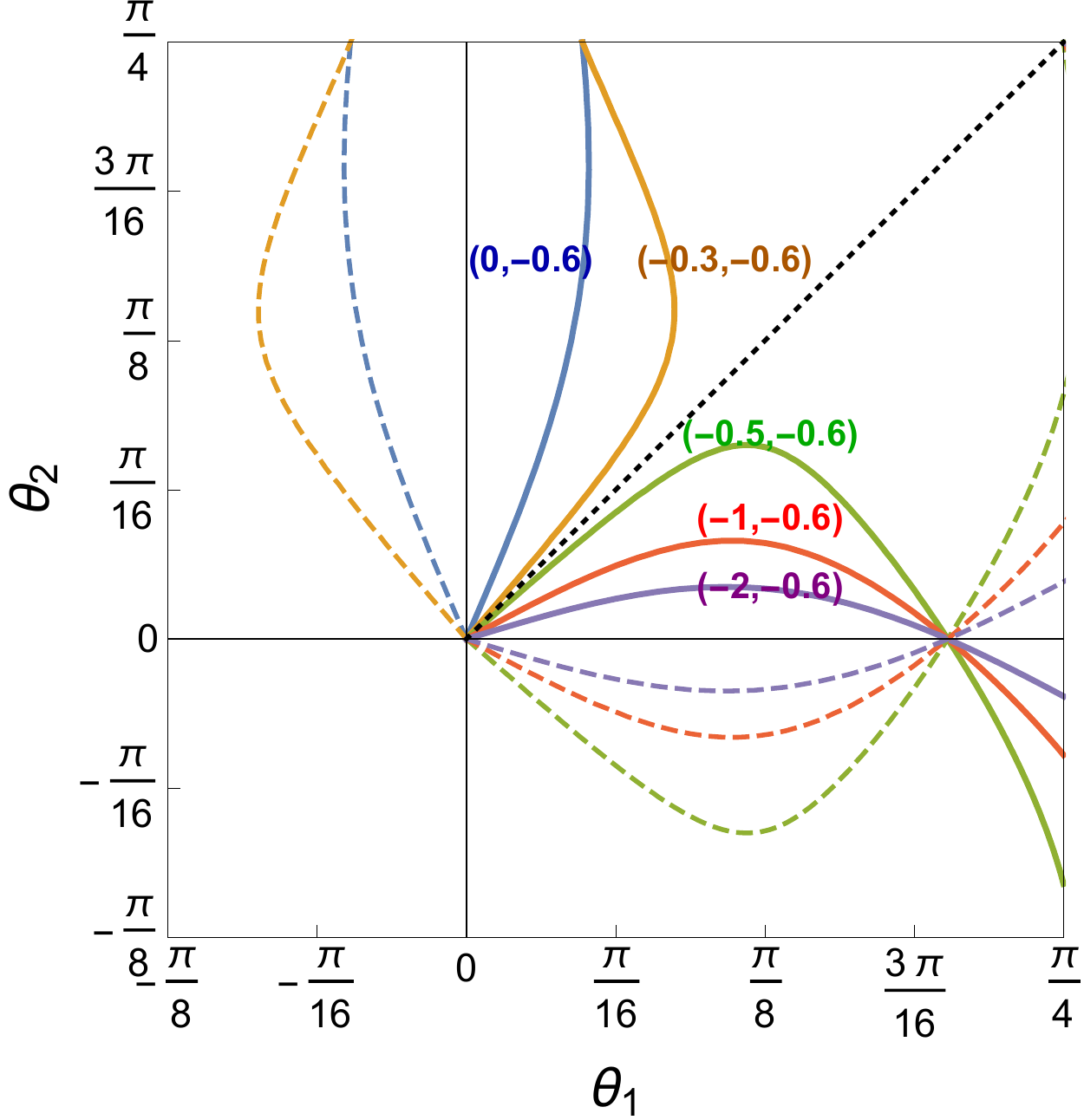}
\includegraphics[width=4.2cm]{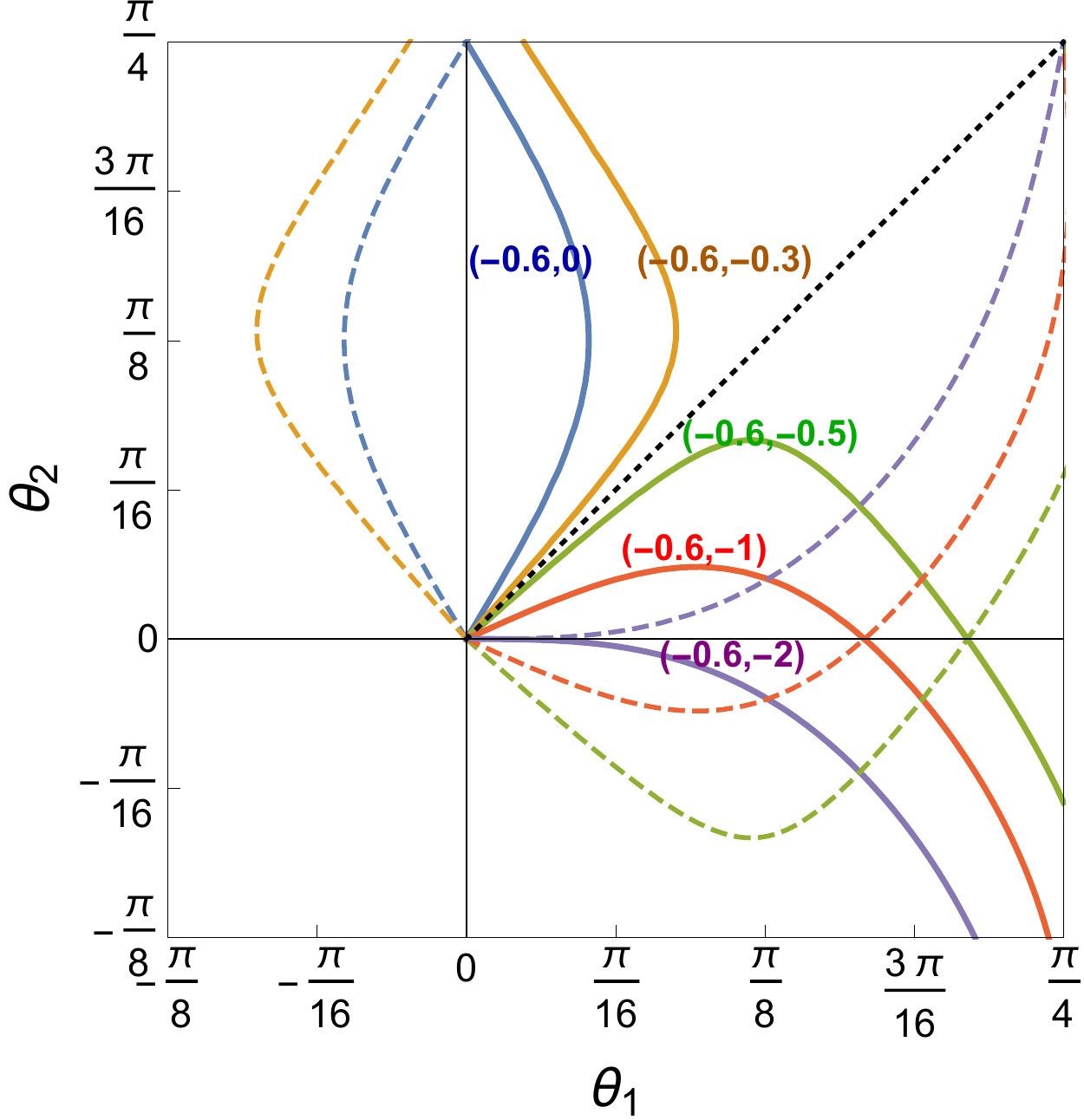}
\caption{
The contour lines shows the relations between $\theta_1$ and $\theta_2$
imposed by two tadpole conditions, denoted by solid and dashed lines respectively, with fixed $\Omega_2 = -0.6$ (left) and fixed $\Omega_1 = -0.6$ (right). 
}
\label{fig:Fig1}
\end{figure}

The radiatively generated scalar potential in Eq.~\ref{eq:loopcorr} further triggers electroweak symmetry breaking
and  induces VEVs for the  GBs $h_{1,2}$ in visible sector. 
The  VEVs of the fields $H_{1,2}$ are parametrized as
\bea
	\langle H_1 \rangle \equiv \left(\begin{array}{c} 0 \\ f_1 \sin \theta_1 \\ 0 \\ f_1 \cos \theta_1 \end{array}\right), \qquad
	\langle H_2 \rangle \equiv \left(\begin{array}{c} 0 \\ f_2 \sin \theta_2 \\ 0 \\ f_2 \cos \theta_2 \end{array}\right).
\eea
where $\theta_1 \equiv \frac{\langle h_{1} \rangle}{f_1}, \quad \theta_2 \equiv \frac{\langle h_{2} \rangle}{f_2}$. 
Similar to 2HDM, $t_\beta \equiv \tan\beta = \frac{f_2}{f_1}$, $\delta_{45} =\delta_4 + \delta_5$ and $\delta_{345} \equiv \delta_3 + \delta_4 + \delta_5$ are used.
Imposing tadpole conditions on Eq.~\ref{eq:loopcorr} determine $\theta_{1,2}$. 
We will neglect $\delta_{6,7}$ terms, because  either $\delta_{6,7} =0 $ in Type-I or $\delta_{6,7} \ll \delta_{1-5}$ ($y_2 \ll g \ll y_1$) in aligned Yukawa 
structure. 
The tadpole conditions are
\bea
&& \sin 4\theta_1 + \Omega_1 \sin 4 \theta_2  +  \Omega_2 \sin 2(\theta_1 + \theta_2) = 0,\nn\\
&& \sin 4\theta_1 - \Omega_1 \sin 4 \theta_2  + \Omega_2  \sin 2(\theta_1 - \theta_2) = 0,
\label{eq:Ccondition}
\eea
where $\Omega_1 = t_\beta^4 \delta_2/\delta_1$ and $\Omega_2 = t_\beta^2 \delta_{345}/\delta_1$.
We are interested in the region $\Omega_{1,2} < 0$ because $\delta_1 >0, \delta_{2-5} <0$.  
If $|\Omega_1 +\Omega_2| >1$ the two conditions are symmetric under $\theta_2 \leftrightarrow -\theta_2$. 
While if $|\Omega_1 +\Omega_2| <1$ they are symmetric under $\theta_1 \leftrightarrow -\theta_1$.
The solutions should be 
\bea
\begin{cases}		
\theta_2 =0,\,\, \theta_1 < \pi/4,  &  \text{for } \, |\Omega_1 +\Omega_2| >1 \\
\theta_1 =0,\,\, \theta_2 > \pi/4,  &  \text{for } \, |\Omega_1 +\Omega_2| <1.
\end{cases}
\label{eq:solutions}
\eea
Thus only one $H_i$ further generates a VEV after radiative symmetry breaking. 
We plot the $(\theta_1, \theta_2)$ contours imposed by tadpole conditions for different $(\Omega_1, \Omega_2)$ in Fig.~\ref{fig:Fig1}.
We note that $\Omega_2$ alone could determine $\theta_{1,2}$ which is intersection point between solid and dashed curves,  
while $\Omega_1$ only controls the convex behavior of the curves.

To obtain the electroweak vacuum $v < f$,  $\theta_i < \pi/4$ is required, which spontaneously breaks the $\mathbb{Z}_2$ symmetry.  
This implies  $|\Omega_1 +\Omega_2| >1$ and $\theta_2 = 0$ in Eq.~\ref{eq:solutions}. 
The electroweak vacuum thus has $v \equiv f_1 \sin\theta_1 = 174$ GeV. 
The tadpole conditions reduce to one condition
\bea
\sin^2\theta_1   = \frac{v^2}{f_1^2} \equiv  \frac12\left( 1 + t_\beta^2\frac{\delta_{345}}{2\delta_1}\right).
\label{eq:sinthf}
\eea
Because of $\delta_1 <0, \delta_{345}>0$, we have $\theta_1 < \pi/4$. 
Furthermore, if $t_\beta$ is large ($f_2 > f_1$), $\theta_1$ could be much smaller than $\pi/4$.
Therefore, $t_\beta$ controls the tuning behavior between $v$ and $f_1$, and it is natural to realize such tuning by setting $f_2 \gg f_1$. 
Let us understand the purely radiative breaking mechanism physically. 
The leading terms in the Higgs potential can be parameterized by 
\bea
	V(h_1) &=& f_1^4 \delta_1 \left[\sin^4(\frac{h_1}{f_1}) + \cos^4(\frac{h_1}{f_1}) \right] + f_1^4 t_\beta^2 \delta_{345} \cos^2(\frac{h_1}{f_1}) \nn\\ 
	  & \simeq &   - (2 + \Omega_2)\delta_1 f_1^2 |h_1|^2 + \frac{8 + \Omega_2}{3} \delta_1 |h_1|^4. 
	\label{eq:higgsh4}
\eea
Both the quadratic and quartic terms are loop-suppressed.  
As mentioned in introduction, if the quadratic term is much smaller than the quartic term, the electroweak VEV could be obtained.
Here let us expand the Higgs quadratic term in Eq.~\ref{eq:higgsh4}:
\bea
	\mu_{h_1}^2 = 2 \delta_1 f_1^2  + \delta_{345} t_\beta f_1^2. 
\eea
Since the Yukawa and gauge corrections have  $\delta_1 > 0$ and $\delta_{345} < 0$ respectively, the Higgs mass squared is 
suppressed by cancellation between Yukawa and gauge corrections.
Note $t_\beta$ plays an important role: only when $t_\beta$ is not so small, cancellation in quadratic term is adequate. 
This implies a moderate tuning $f_1 < f_2$, which induces tuning between Yukawa and gauge corrections  correspondingly.
As a measure of the naturalness, the estimation of the fine-tuning is
\bea
	\Delta = \left|\frac{2\delta m^2}{m_h^2}\right|^{-1}  \simeq \left|\frac{3 y_t^2 m_{t_B}^2}{4\pi^2 m_h^2}\right|^{-1} \sim \frac{2v^2}{f_1^2}.
	\label{eq:finetuning}
\eea
Unlike the soft breaking or tadpole breaking mechanism, the tuning is realized via hierarchy between $f_1$ and $f_2$. 
From Eq.~\ref{eq:sinthf}, for a level of tuning 10\%, $t_\beta \simeq 3$ ($f_2 \simeq 3 f_1$).

%

\begin{figure}[!t]
\begin{center}
\includegraphics[width=4.2cm]{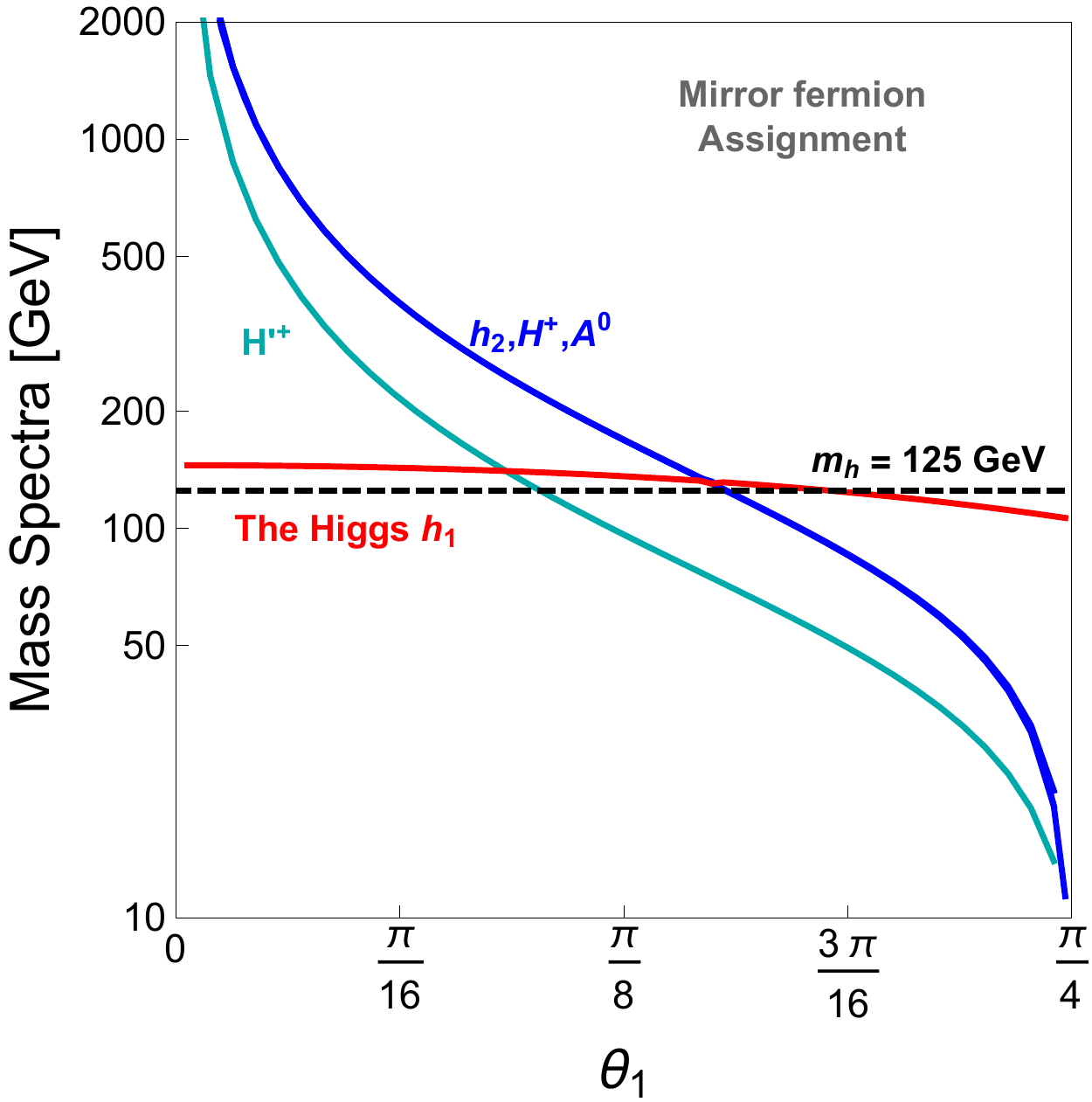}
\includegraphics[width=4.2cm]{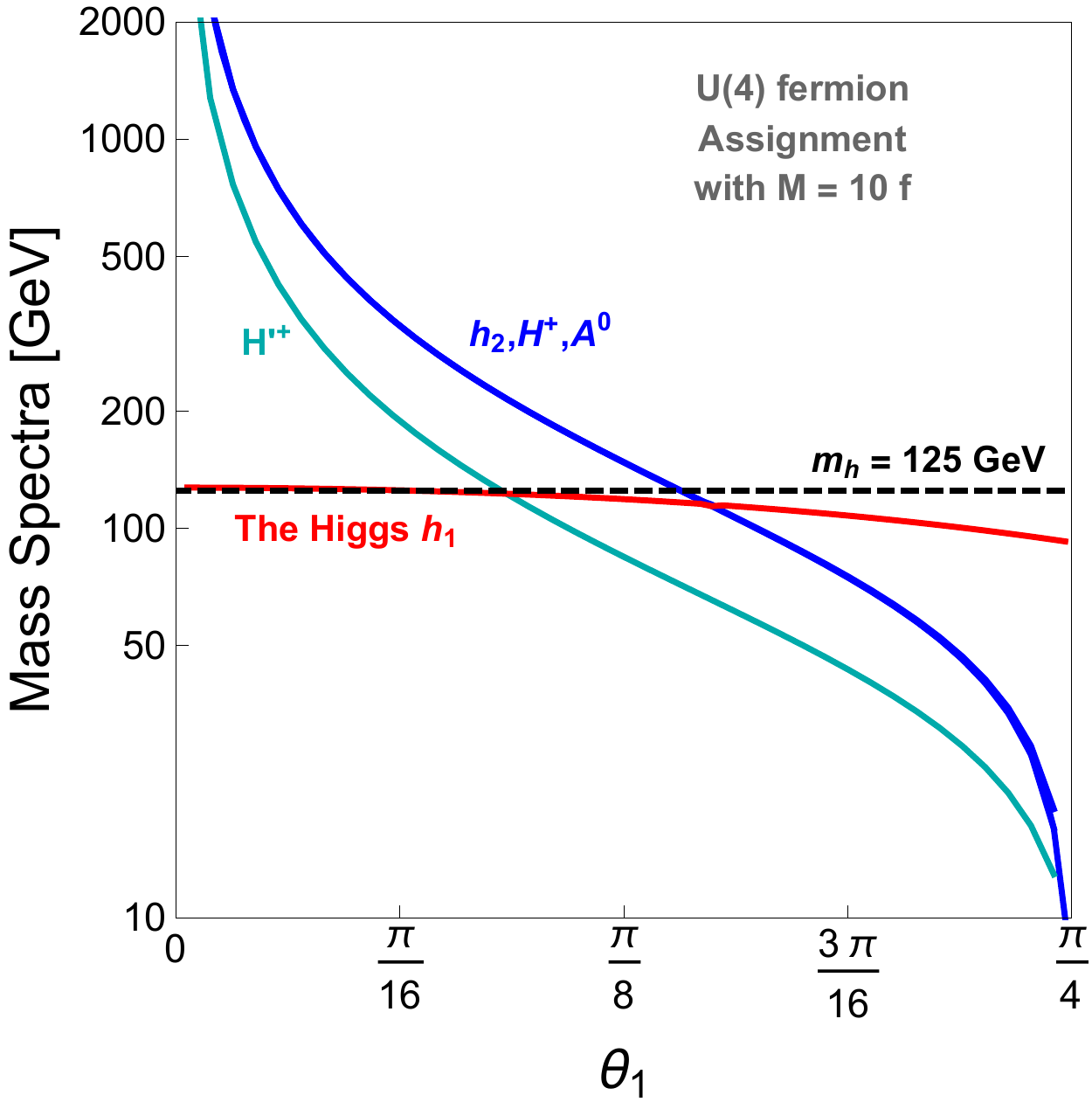}
\end{center}
\caption{
The mass spectra of the PGBs as function of $\theta_1$  for ``mirror fermion" (left panel) and ``$U(4)$ fermion" (right panel, $M = 10 f$ is taken) assignments. 
}
\label{fig:figmas}
\end{figure}

The purely radiative symmetry breaking only generate VEV $\langle H_{1A} \rangle$, but not $\langle H_{2A} \rangle$. 
Zero $H_{2A}$ VEV  ($\theta_2 = 0$) implies that the second Higgs $H_{2A}$ in $A$ sector is inert Higgs doublet~\cite{Barbieri:2006dq}, 
which does not mix with $H_{1A}$. 
In visible sector, particles in $H_{iA}$ are identified as GBs $\bm{h}_i$. 
Among them, ($z^{0,\pm}$) in $H_{1A}$ and $(H^\pm, A^0)$ in $H_{2A}$ have
\bea
	m_{z^0}^2  &=& m_{z^\pm}^2 = 0, \quad \textrm{(exact GBs eaten by $W_A, Z_A$)}, \nn\\
	m_{H^\pm}^2 & \simeq &  2 \delta_1 f_1^2 t_\beta^2 \cos^2 2\theta_1 - 2 \delta_2 f_2^2 
	- \delta_{45}f_1^2\sin^2 \theta_1, \nn\\
	m_{A^0}^2 & \simeq &  
      2 \delta_1 f_1^2 t_\beta^2 \cos^2 2\theta_1 - 2 \delta_2 f_2^2 
     	- 2\delta_{5}f_1^2\sin^2 \theta_1,
	\label{eq:inertmass}
\eea 
And two CP-even GBs $h_1$ in $H_{1A}$ and $h_2$ in $H_{2A}$, which do not mix together due to zero $H_{2A}$ VEV, have
\bea
	m_{h_1}^2 & \simeq &  
	8 \delta_1 f_1^2 \sin^2\theta_1  , \nn\\
	m_{h_2}^2 & \simeq & - 2 \delta_2 f_2^2 +  2 \delta_1  f_1^2 \cos^22\theta_1/t_\beta. 
\eea
Here $h_1$ is identified as the SM Higgs boson.
However, in twin sector $B$, the GBs in $H_{1B}$ and $H_{2B}$ are mixed due to the VEVs $f_{1,2}$. 
The rotation angle $\beta_B$ between $C^{\pm}_1 (N^0_1)$  and $C^{\pm}_2 (N^0_2)$ is defined as $\tan \beta_B = t_\beta/\cos\theta_1$. 
Performing rotation to mass basis, we obtain 
\bea
	m_{N^0}^2 &=& m_{C^\pm}^2 = 0, \quad \textrm{(exact GBs eaten by $W_B, Z_B$)}, \nn\\
		m_{H'^\pm}^2 & = & -\left(\delta_{45} +  \delta_7 \tan \beta_B\right) f_1^2( \cos^2\theta_1 + t_\beta^2), \nn\\
	m_{A'^0}^2 & = & -\left(2\delta_{5} + \delta_7 \tan \beta_B\right) f_1^2 \frac{(\cos^2\theta_1 + t_\beta^2)^2}{\cos^2\theta_1}.
\eea
%
Fig.~\ref{fig:figmas} shows mass spectra of the pGBs in two cases. 
In ``mirror fermion" case, the pGB masses only depend on single parameter $\theta_1$. 
Thus the requirement of a 125 GeV Higgs mass determines $\theta_1 = 0.57$, which corresponds to $t_\beta = 2$. 
In ``$U(4)$ fermion" case, mass spectra depend on both $\theta_1$ and vectorlike fermion $\tilde{q}_A$ mass $M$, which should have $M \le 4\pi f$. 
As the $M$ takes smaller value than $4\pi f$, the  $\theta_1$, obtained from the 125 GeV Higgs mass condition, gets smaller value.   
However, when $M = 8 f$, $\theta_1$ reaches zero, which put a lower cutoff for $M$. 
In the following, we take $M = 10 f$ as the benchmark point.

The current limits on NP searches at the LHC put very strong constraints on new particles. 
New particles in $A$ sector are the inert Higgses $H^\pm, h_2, A^0$, which is typically tightly constrained~\cite{Barbieri:2006dq}. 
However, if the inert Higgses have nearly degenerate masses,  
it happens to be very difficult to probe this compressed parameter region at the LHC, which has been studied in Ref.~\cite{Blinov:2015qva}.
%
%
For particles in twin sector, it is harder to directly probe them due to zero SM charges. 
However, because of twin colorness, there are rich twin hadron phenomenology, which has been discussed in Ref.~\cite{Craig:2015xla, Cheng:2015buv}.  
For simplicity, in the following we adopt minimal twin matters: fraternal twin Higgs~\cite{Craig:2015pha}, in which only the third generation twin fermions are introduced, and  typically twin lepton is identified as dark matter candidate. 
In this scenario,  
the twin photon and $A'^0$ could be either massless or massive depending on gauge and fermion assignments.
For example, the aligned Yukawa structure could lift the $A'^0$ mass from zero value. 
If they are massless, they should contribute to dark radiation.
Depending on temperature of thermal decoupling between visible and twin sector~\cite{Craig:2015xla}, 
the number of effective neutrino species $\Delta N_{\rm eff}$ could be adjusted to be within the range of recent Planck measurement $0.11 \pm 0.23$~\cite{Ade:2015xua}. 
We leave the detailed discussion in future study~\cite{Yufuture}.

\begin{figure}[!t]
\begin{center}
\includegraphics[width=4.1cm]{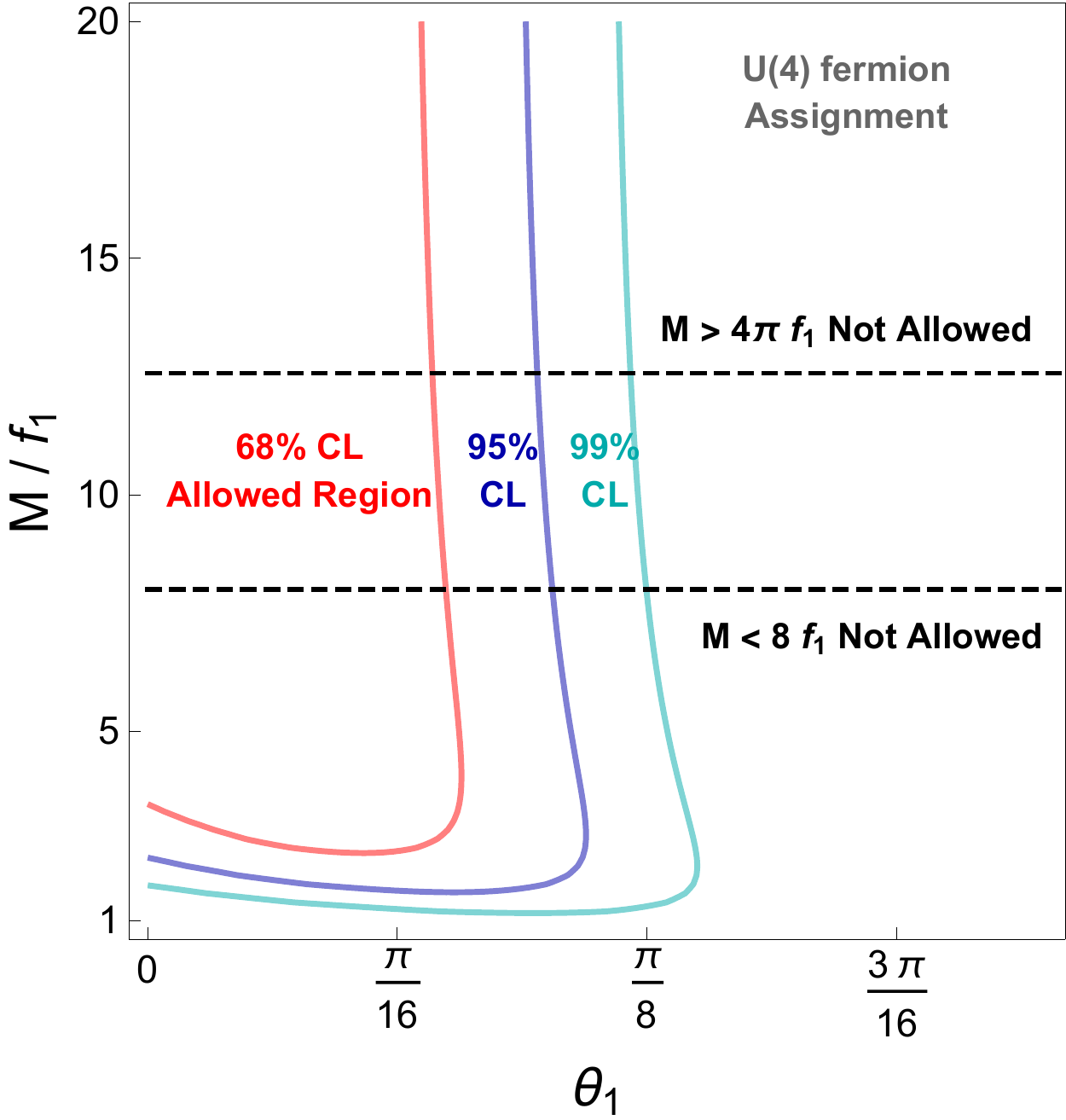}
\includegraphics[width=4.2cm]{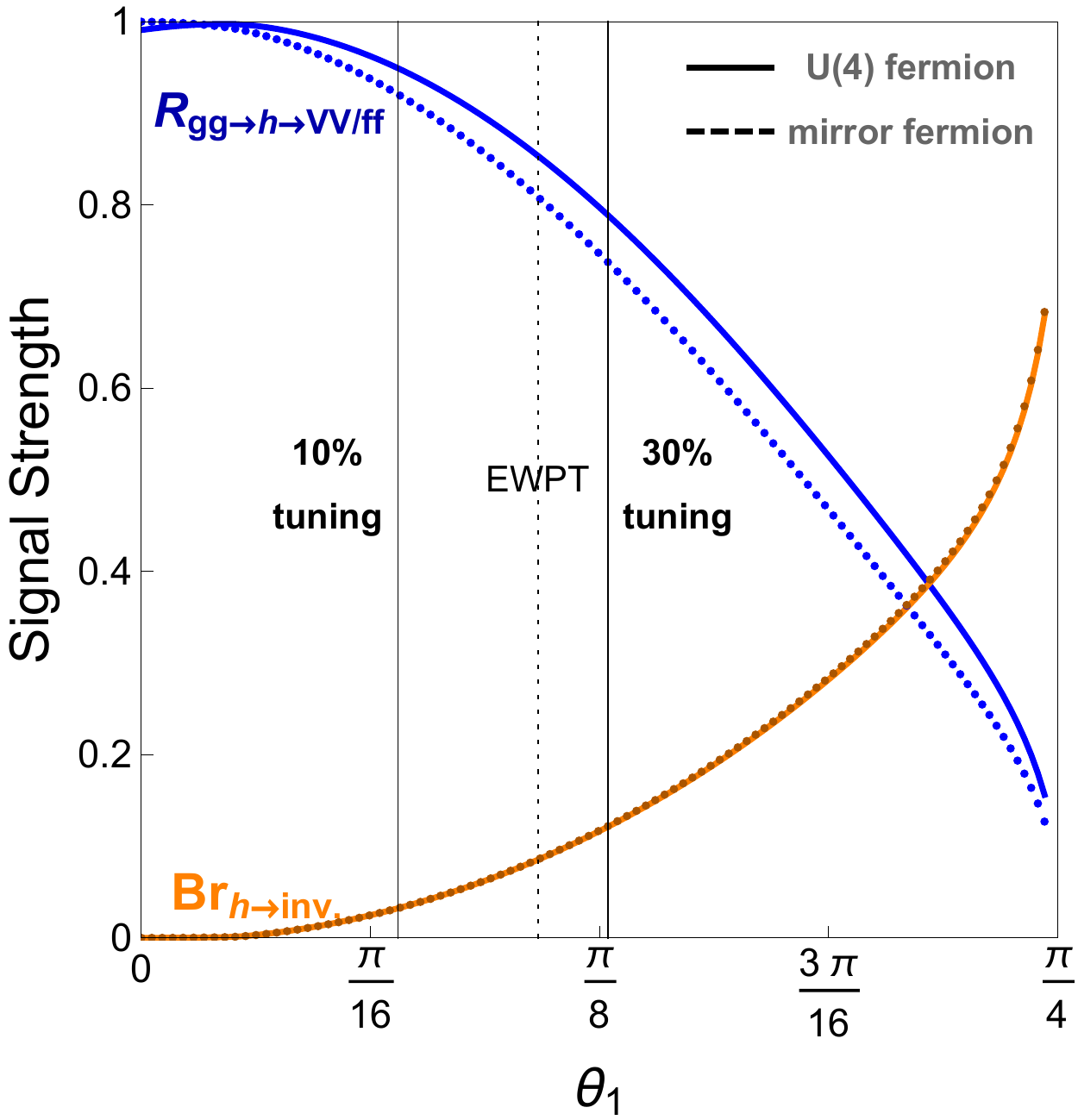}
\end{center}
\caption{
(Left) the allowed contours on $(\theta_1, M/f_1)$ at 68\%, 95\%, 99\% CLs in  ``$U(4)$ fermion" assignment.
(Right) signal strength in gluon fusion channel (blue)
and invisible branching ratio (orange) as function of $\theta_1$ in ``$U(4)$" (solid) and ``mirror" (dotted) fermions.
}
\label{fig:figratio}
\end{figure}

The measured Higgs production and decay cross sections, and the upper limits on Higgs invisible decays at the LHC~\cite{Aad:2015gba}
also provide strong constraints on model parameters. 
The tree-level couplings of the Higgs boson to fermions and bosons in $A$($B$) sector are altered by a factor $\cos\theta_1$ ($\sin\theta_1$) 
relative to SM. 
We assume masses of twin particles are altered by a factor $\cot\theta_1$ relative to SM.  
In  ``$U(4)$ fermion" case, there are also a heavier vectorlike top $T$ which mixes with the top quark through mixing angle $\cos\theta_R = y_t f/\sqrt{M^2 + y^2f^2}$. 
This modifies the top-Higgs coupling to $y_t \cos\theta_1 \cos\theta_R$, and a new $TTh$ coupling has $y_t \sin\theta_1 \sin\theta_R$.   
We calculate various Higgs signal strengths $\mu_{pp\to h_1 \to ii} =  \sigma(pp\to h_1) \textrm{Br}_{h_1 \to ii}/\sigma_{\rm SM}\textrm{Br}_{\rm SM}$, and invisible decay width. 
Based on Higgs signal strengths at the 8 TeV LHC with 20.7 fb$^{-1}$ data~\cite{Aad:2015gba}, 
we perform a global fit on model parameters~\cite{Yufuture}. 
Fig.~\ref{fig:figratio} (left) shows the allowed contours on $(\theta_1, M/f_1)$ at 68\%, 95\%, 99\% confidence levels (CLs) in  ``$U(4)$ fermion" case. 
Fig.~\ref{fig:figratio} (right) plots the signal strength of gluon fusion $gg\to h_1 \to VV/ff$, and invisible decay width in two assignments. 
%
%
We list our global $\chi^2$-fitting results on Higgs signal strengths in ``mirror fermion" (``$U(4)$ fermion" with $M = 10 f$) assignments:
\bea
	\theta_1 \equiv \frac{v}{f_1} < 0.25 \,\,(0.31) \,\,\,\,{\textrm{@ 95\% CL}}.
\eea
This limit rules out the whole parameter region of the ``mirror fermion" case, but ``$U(4)$ fermion" case is still viable.
The electroweak precision test put additional constraints on model parameters. 
The contribution from inert Higgs doublet is negligible due to their degenerated masses. 
Thus the dominant Logarithmic contributions to $S$ and $T$ parameters~\cite{Burdman:2014zta} are $\alpha T(\alpha S) \sim \mp \sin^2\theta_1\log\frac{\Lambda}{m_Z}$. 
We also estimate the levels of tuning of about 10\% and 30\%, which are shown in Fig.~\ref{fig:figratio} (right). 
The high luminosity LHC will improve sensitivity of signal strengths to around 5\% assuming current uncertainty with  3 ab$^{-1}$ luminosity~\cite{atlas:HLLHC}.
This indicate that we could probe this model with about 12\% tuning by the end of high luminosity LHC run.  
%



In summary, we have investigated a minimal two twin Higgs model, in which the Higgs boson is a pseudo-Goldstone boson after symmetry breaking  $[U(4)\times U(4)] \to [U(3)\times U(3)]$. 
The $\mathbb{Z}_2$ symmetry, which protects the Higgs mass against quadratic divergence, is spontaneously broken by radiatively generated Higgs potential. 
The vacuum misalignment $v < f$ is realized radiatively via cancellation of gauge and Yukawa corrections to the Higgs mass term. 
This minimal setup for spontaneous $\mathbb{Z}_2$ breaking has less parameters and it has predictive but rich phenomenology.

\noindent {\textbf{Acknowledgements}} - The author thanks Can Kilic and Nathaniel Craig for valuable discussions. This work was supported by DOE Grant DE-SC0011095.



\end{document}